\documentclass[useAMS,usenatbib,usegraphicx]{mn2e}
\usepackage{graphicx,amssymb}
\usepackage{color}

\title[Photometric study of the Blazhko star {RZ~Lyr}]{An extensive 
photometric study of the Blazhko RR Lyrae star RZ~Lyr\thanks{Based on observations collected primarily with the automatic 60-cm telescope of Konkoly Observatory, Budapest, Sv\'abhegy}}
\author[Jurcsik et al.]{J. Jurcsik$^{1}$, \'A. S\'odor$^{1}$, G. Hajdu$^{1}$, B. Szeidl$^{1}$, \'A. D\'ozsa$^{2}$, K. Posztob\'anyi$^{3}$, P. Smitola$^{4}$ \and B. Belucz$^{4}$, V. Feh\'er$^{4}$, Zs. K\H ov\'ari$^{1}$, L. Kriskovics$^{1}$, E. Kun$^{2}$, L. Moln\'ar$^{1}$, I. Nagy$^{4}$ \and K. Vida$^{1}$, N. G\"or\"og $^{4}$
\\
$^{1}$Konkoly Observatory of the Hungarian Academy of Sciences, H-1525 Budapest PO Box 67, Hungary\\
$^{2}$Department of Experimental Physics and Astronomical Observatory, University of Szeged, 6720 Szeged, D\'om t\'er 9, Hungary\\
$^{3}$Visiting observer at Konkoly Observatory\\
$^{4}$E\"otv\"os University, Dept. of Astronomy, H-1518 Budapest PO Box 49, Hungary
}
\begin{document}

\date{Accepted 2011  Received 2011 Aug; in original form 2011 Dec}

\pagerange{\pageref{firstpage}--\pageref{lastpage}} \pubyear{2012}

\maketitle

\label{firstpage}
\begin{abstract}

The analysis of recent, extended multicolour CCD and archive photoelectric, photographic and visual observations has revealed several important properties of RZ~Lyr, an  RRab-type variable exhibiting large-amplitude Blazhko modulation. On the time-base of $\sim$110 yr, a strict anticorrelation between the pulsation and modulation period changes is established. The light curve of RZ Lyr shows a remarkable bump on the descending branch in the small-amplitude phase of the modulation,  similarly to the light curves of bump Cepheids. We  speculate that the stellar structure temporally suits a 4:1 resonance between the periods of the fundamental and one of the higher-order radial modes in this modulation phase. The light-curve variation of RZ~Lyr can be correctly fitted with a two-modulation-component solution; the 121 d period of the main modulation is nearly but  not exactly four times longer than the period of the secondary  modulation component. Using the inverse photometric method, the variations in the pulsation-averaged values of the physical parameters in different phases of both modulation components are determined.

\end{abstract}
\begin{keywords}
stars: horizontal branch -- 
stars: oscillations -- 
stars: variables: RR Lyr -- 
stars: individual: RZ~Lyr --
techniques: photometric -- 
methods: data analysis 
\end{keywords}

\section{Introduction}

The light-curve modulation phenomenon of RR Lyrae stars, the so-called Blazhko effect, is still one of the unsolved problems of stellar pulsation theory. Despite the recent ground-based and satellite observational \citep{ju08,mw2,bl,m5bl,rrg2,czl,ch10,gu11,ko11} and theoretical \citep{st1,st2,st3,sm,pd,pdd} efforts aiming at disclosing the root cause of the light-curve changes, no convincing explanation for the phenomenon has been given yet.

We have learned a lot about the complexity of the frequency spectra of the light variation, the occurrence rate of the modulation, the long-term changes of the modulation properties and the changes in the mean global physical parameters during the modulation cycle in recent years, but none of the proposed explanations is capable to elucidate all the observed features of these peculiar variable stars. Therefore, we think
that further detailed studies may still help in understanding the Blazhko phenomenon.

An ideal target for studying the Blazhko effect is RZ Lyr  ($<V>=11.7$ mag, P=0.511 d, $\alpha_{2000} = 18^{\rm h}43^{\rm m}37.\!\!^{\rm s}9$, $\delta_{2000} = +32{\degr}47'53.\!\!^{\prime\prime}9$). Its photometric history dates back to the beginning of the 20th century. 
Since its discovery by \cite{wi03}, the star has been the target of a large number of visual \citep{ba51,ba52,ts53,ts58,kl58,be69,ro69,bo72,zv79} and some photoelectric studies \citep{fi66,st66,ro73,bu82}. 
In addition, unprocessed, archive photographic and photoelectric observations of RZ Lyr are also available at the Konkoly Observatory.

The large-amplitude, relatively long-period ($\sim$116 d) Blazhko modulation of RZ Lyr was already established by \cite{ts53}. Based on the quasi-monotonic period decrease of the star \citep[$\alpha= P^{-1} \mathrm{d} P/ \mathrm{d} t=-0.367$ Myr$^{-1}$,][]{lb}, it could be supposed that the light-curve modulation of RZ Lyr does not show an erratic/complex behaviour either.
Therefore, an extended CCD observation series of RZ Lyr was initiated
in 2010 in order to follow the connection between the pulsation and modulation period changes of the star, and to disclose the changes in the mean values of the physical parameters during its Blazhko modulation.

\section{Observations of RZ~Lyrae}

The analysis of the pulsation and modulation properties of RZ~Lyr is based mostly on a recent, extended CCD data set and on previously unpublished, archive photographic and photoelectric observations obtained with the telescopes of the Konkoly Observatory. In addition, all the published photometric information available on the light variation of RZ~Lyr have been collected and are utilized.

\subsection{Konkoly observations}
\label{konkoly}
\begin{figure}
\centering
\includegraphics[width=8.1 cm]{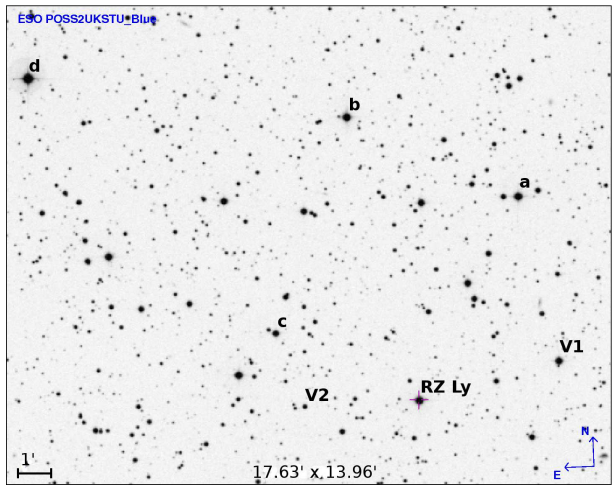}
\caption{Identification of the comparison stars and the new variables in the field of view of RZ~Lyr. }
\label{map}
\end{figure}

A total number of about 1400 photographic exposures were taken with the 16-cm astrograph of the Konkoly Observatory at Budapest between 1950 and 1954, on 130 plates. The plates were digitalized using an EPSON Perfection V750 flatbed transparency scanner. The photographic densities of RZ~Lyr and the comparison stars were determined using standard IRAF\footnote{{\sc IRAF} is distributed by the National Optical Astronomy Observatories, which are operated by the Association of Universities for Research in Astronomy, Inc., under cooperative agreement with the National Science Foundation.}
aperture photometry packages. The magnitudes of RZ~Lyr were evaluated using Tycho-2 $B$ magnitudes \citep{tycho} of the surrounding stars.

Photoelectric  observations of RZ~Lyr were collected with the 60-cm telescope of the Konkoly Observatory in 1958, 1959 with RCA 1P21, and in 1968, 1969 and 1972 with EMI 9052 B photomultiplier tubes. About 700 observations were obtained in $B$ and $V$ bands. Three different comparison stars were used during the observing runs. Standard magnitudes of the comparison stars are taken from the comparison sequences of the AAVSO Variable Star Plotter\footnote{http://www.aavso.org/vsp/} (5671dpv) and from \citet{st66}. The $BVI_\mathrm{C}$ magnitudes of the comparison stars are summarized in Table~\ref{comp}.

CCD observations of RZ~Lyr were obtained on 109 nights between 2010 April and 2011 September with the automatized 60-cm telescope of the Konkoly Observatory at Sv\'abhegy, Budapest, equipped with a Wright Instruments 750 $\times$ 1100 pixel CCD camera and $BVI_\mathrm{C}$ filters. About 2300 exposures were taken in each band. Relative magnitudes of RZ~Lyr were measured against the comparison star `a'. The photoelectric and CCD observations were transformed to Johnson--Cousins $BVI_\mathrm{C}$ magnitudes by standard procedures.

Originally, another blue star (2MASS 18431843 +3248587) was selected as comparison for the CCD data, which, however, proved to be  variable (V1). Without a  systematic variable search, another new variable (V2: 2MASS 18435377 +3247459) was also identified in the field. V1 and V2 were measured relative to the comparison stars `a' and `c', respectively. The light curves of the new variables are given in the Appendix.
The comparison stars and the new variables (V1,V2) are identified in Fig.~\ref{map}. 
The Konkoly photometric observations and maximum times of RZ~Lyr are available as Supporting Information with the electronic version of this paper.
Tables~\ref{data} and~\ref{maxima} give samples regarding their form and content.

\begin{table}
\caption{Johnson-Cousins magnitudes of the photoelectric and CCD comparison stars}
 \label{comp}
  \begin{tabular}{@{\hspace{-2pt}}l@{\hspace{0pt}}c@{\hspace{3pt}}c@{\hspace{3pt}}c@{\hspace{3pt}}l}
  \hline
Comp.star/2MASS ID&$B$&$V$&$I_\mathrm{C}$&Ref.\\
\hline
a 18432371+3253475 &11.546(19) &11.055(11)   &10.454(29)&AAVSO VSP$^{a}$\\
b 18434732+3256105 &11.440(20) &10.814(10)   &10.120(21)&AAVSO VSP$^{a}$\\
c 18435761+3249546 &12.846(16) &11.977(13)   &11.009(18)&AAVSO VSP$^{a}$\\
d 18443148+3257263 &9.966  &9.800    &--&~\citet{st66} \\
\hline
\multicolumn{5}{l}{\footnotesize{$^{a}$ http://www.aavso.org/vsp/}}\\
\end{tabular}
\end{table}

\begin{table}
\caption{Konkoly photometric observations of RZ~Lyr. The full table is available as Supporting Information with the online version of this paper.}
 \label{data}
  \begin{tabular}{lccccc}
  \hline
HJD        &mag    &Detector        &Filter    &Comparison\\
2400000+   &           &            &          &star\\
\hline
55294.64087&12.369     &CCD         &$B$       &\\
55294.64436&12.339     &CCD         &$B$       &\\
...&...&...&...& &\\
55294.64238&11.942     &CCD         &$V$       &\\
...&...&...&...& &\\
55294.64341&11.325     &CCD         &$I$       &\\
...&...&...&...& &\\
36413.4595 &12.656     &pe          &$B$       &a\\
...&...&...&...&...&\\
36413.4643 &12.145     &pe          &$V$       &a\\
...&...&...&...&...&\\
33369.501  &12.29      &pg          &          &\\
        ...&...        &...         &          &\\
\hline
\end{tabular}
\end{table}

\begin{table}
\caption{Times and brightnesses of maxima  redetermined from published visual and photoelectric light curves and determined from the Konkoly observations. The full table is available as Supporting Information with the online version of this paper.}
 \label{maxima}
  \begin{tabular}{lllll}
  \hline
Times of maxima&$pg/B$&$vis/V$&$I_\mathrm{C}$&references\\
HJD-2400000&mag&mag&mag&\\
\hline
\hline
26074.5430 & &10.67& & 1\\
26209.4900 & &11.09& & 1\\
...&...&...\\
33369.541   &11.04&&&3 \\
33372.610   &11.06&&&3\\
...&...\\
36413.513      &10.98   &10.81&&3\\
...&...&...\\
55346.356      &10.983  &10.842  &10.617&3\\
...&...&...&...\\
\hline
\multicolumn{5}{l}{\footnotesize{$^{*}$ reference numbers are the same as in Table~\ref{p-p}}}.
\end{tabular}
\end{table}

\subsection{Other photometric data }
\label{archiveoc}

In the GEOS\footnote{http://rr-lyr.ast.obs-mip.fr/}  \citep{lb} data base, all the published maximum times have been collected and many other maximum observations are also given.  We have inspected all the GEOS data by  checking the given maximum timings  against the original light-curve data, if they are available
\citep{be69,fi66,kl58,m,ro69,ts53,ts58,ts69,zv79}.
Additionally, multiple entries are corrected and outlying points are removed.

Besides maximum timings, light-curve data and maximum-brightness values are also used to detect changes in the pulsation and modulation properties. In Table~\ref{maxima}, maximum times and brightness values determined and/or redetermined from the original light-curve data published in \cite{ts53,kl58,fi66,zv79}, which were utilized for determining the modulation periods (see Table~\ref{p-p}) are also given.

No good quality light curve of RZ Lyr was published between 1972 and 2010. However, thanks to the efforts of mainly the GEOS\footnote{http://geos.webs.upv.es/}  and BAV\footnote{http://www.bav-astro.de/}  observers the maximum times available in the GEOS data base cover this period \citep{a1,a2,a3,a4,h1,h2,h3,h4,h5,h6,h7,l1,l2,l3,l4,l5,l6,l7,l8,l9,l10,l11,l12,s1,s2}.

\section{Long-term changes in the pulsation and the Blazhko periods}

\begin{figure}
\begin{centering}
\includegraphics[width=8.8 cm]{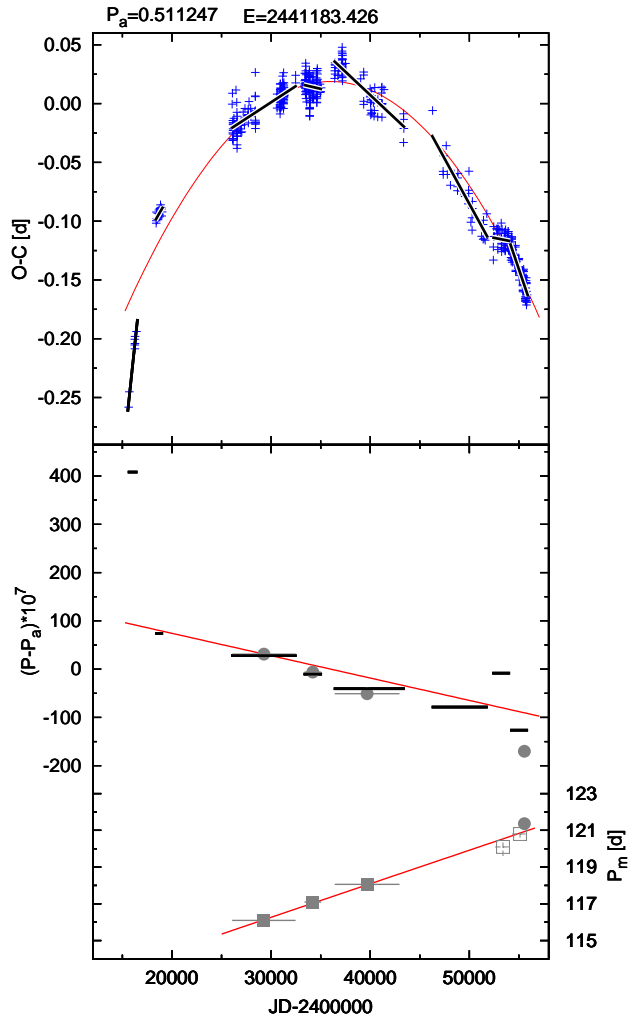}
\caption{Top panel: $O-C$ values of the maximum-timing observations of RZ~Lyr. The $O-C$ data reflect a quasi-monotonic period decrease during the $\sim$110 yr of the observations; a parabolic fit to the $O-C$ data is also shown. Different segments of the data are fitted with lines for a direct determination of the temporal period values. Bottom panel: pulsation and modulation periods calculated for different segments of the data. The temporal pulsation periods that correspond to the $O-C$ data segments are indicated by horizontal black lines, gray symbols show the pulsation periods determined from the light-curve data. Modulation periods determined from maximum brightness,  phase and light-curve analyses are plotted by filled and empty squares and filled circles, respectively. The linear pulsation and modulation period variations shown by straight lines correspond to the parabolic $O-C$ fit and the linear fit of the Blazhko period values. }
\label{oc}
\end{centering}
\end{figure}

\begin{table*}
\caption{Pulsation and Blazhko periods of RZ~Lyr at different epochs}
 \label{p-p}
  \begin{tabular}{clrrlllll}
  \hline
Time-interval&${P}_{0}$ &Error&Diff$^{*}$ & ${P}_\mathrm{m}$ &Error&N$_{\mathrm{max}}$&Method$^{**}$& Data (ref)$^{***}$ \\
JD$-2400000$&d&\multicolumn{2}{c}{$10^{-7}$ d}&d&d&&&\\
\hline
26074 -- 32473 &0.5112501 &1 &0&116.1    &0.1&88 & a& visual (1), (2)\\
33369 -- 35041 &0.5112464 &4 &-14&117.1  &0.2&53 & a& photographic (3)\\
36413 -- 42955 &0.5112419 &2 &-34&118.04 &0.04&34& a& photoelectric V (3), (4), (5), (6), (7)\\
52416 -- 54014 &0.5112462 &10& 72 &120.1 &0.5&48 & b& maximum timings (8)\\
54205 -- 55826 &0.5112343 &7 &-39 &120.8 &0.2&85 & b& maximum timings (3), (8)\\
55294 -- 55826 &0.5112300 &1 &-80 &121.36&0.04&  & c& CCD (3)\\
\hline
\multicolumn{9}{l}{\footnotesize{$^{*}$ difference from the period value predicted from a parabolic $O-C$ fit }}\\
\multicolumn{9}{l}{\footnotesize{$^{**}$ a: pulsation and modulation periods from fitting light curve and maximum brightness data, respectively}}\\
\multicolumn{9}{l}{\footnotesize{\,\,\,\,\, b: pulsation and modulation periods from fitting maximum timing data}}\\
\multicolumn{9}{l}{\footnotesize{\,\,\,\,\, c: pulsation and modulation periods from complete light curve analysis}}\\
\multicolumn{9}{l}{\footnotesize{$^{***}$ (1) \citet{ts53}; (2) \citet{zv79}; (3) this paper; (4) \citet{st66}; (5) \citet{fi66};}}\\
\multicolumn{9}{l}{\footnotesize{\,\,\,\,\,\, (6) \citet{bu82}; (7) \citet{ro73}; (8) GEOS maximum timings }}\\
\end{tabular}
\end{table*}

To follow the pulsation-period changes of RZ~Lyrae, the $O-C$ diagram is constructed using the maximum timings discussed in Section~\ref{archiveoc} complemented with the maximum times obtained from the Konkoly data (Table~\ref{maxima}). According to the $O-C$ variation shown in Fig.~\ref{oc}, the pulsation period has been decreasing more or less continuously during the $\sim$110~yr covered by the observations. A parabolic fit to the $O-C$ data yields an $\alpha = P^{-1}\mathrm{d}P/\mathrm{d}t= -8.85 \times~10^{-10}$~d$^{-1} (-0.323$ Myr$^{-1})$ period-decrease rate.

However, the temporal periods defined by linear fits to the different segments of the $O-C$ data indicate that the period change has not been strictly linear. Although differences between the actually observed and  predicted period values  may arise from an incomplete averaging of the phase modulation of the Blazhko effect in the case of scarce data, this is not the case for the latest epochs. The maximum times available for the time intervals JD 2\,452\,416--2\,454\,014 and JD 2\,454\,205--2\,455\,826 give complete phase coverage of  the 120-d modulation cycle. The high accuracy of the  maximum timings collected by the GEOS team makes it also possible to detect the $\sim$0.02 d phase modulation in these data sets. Therefore, besides the temporal pulsation periods, the Blazhko periods are also derived for these epochs from maximum-timing data.

The archive light curves and maximum-brightness values enable us to determine both the 
pulsation and the Blazhko periods with high accuracy in a homogeneous manner for three epochs. In order to get the most reliable results, only the best-quality, most accurate and homogeneous light curves and maximum-brightness data obtained at the different time-intervals are utilized for this purpose. The pulsation periods are determined by fitting 6th--12th-order Fourier series to the light-curve data, without taking the modulation components into account. The incomplete phase coverage of these data does not make a complete analysis possible. The Blazhko periods are obtained from sinusoidal fits to the maximum-brightness values.

The first data set, which covers $\sim$18 yr, comprises the visual observations of \citet{ts53} and \citet{zv79}. There is no significant systematic difference between the magnitude scales of the two authors. These are the only available photometric information on the light variation of RZ Lyr for this period. Although a lot of visual observations are also available for the second half of the 20th century, the inhomogeneous quality and ill-defined magnitude scales do not render their use in the analysis possible. The pulsation and modulation periods  are determined from the Konkoly photographic data for the 1950--1954 interval. The third data set covering the 1957--1973 period includes the photoelectric observations published in this paper complemented with the photometric data of \citet{st66}, \citet{fi66} and \citet{bu82}. The timings and magnitudes of maximum light published by \citet{ro73} are also utilized to derive the Blazhko period at this epoch.

The recent CCD observations provide accurate pulsation and modulation periods from detailed light-curve analysis as discussed in Section~5.

Table~\ref{p-p} summarizes the pulsation and modulation periods derived for the different data sets using different methods. 

In the bottom panel of Fig.~\ref{oc}, the pulsation and modulation period changes are shown according to the data given in Table~\ref{p-p}. A strong anticorrelation between the pulsation and modulation period changes is detected; while the pulsation period decreased by $\sim$0.000012~d the modulation period increased by $\sim$5~d.
A ${\mathrm d}P_{0}/{\mathrm d} P_\mathrm{m}=-2.4\times 10^{-6}$  period-change ratio is determined. While the modulation period values fit a linear period increase quite accurately, the pulsation period  shows some erratic changes superimposed on the linear period decrease, especially at the latest epochs.

In \cite{m5bl}, period-change data of Blazhko stars have been collected for all the variables with pulsation and modulation periods known at least for two different epochs. The majority of these stars (10 out of 13) show anticorrelated pulsation and modulation period changes while parallel changes in the periods are detected only in three cases (RR Gem, XZ~Dra, V56/M5). All the variables that have enough data to determine both pulsation and modulation periods for more than two epochs show, however, complex period-change behaviour. The correlation/anticorrelation between the pulsation and modulation periods manifests itself, in fact, as a tendency and not as a strict relation in these stars. Such a strict connection between the changes in the pulsation and modulation periods as observed in RZ~Lyr have not been detected in any Blazhko star previously.

\begin{figure*}
\includegraphics[width=17.2cm]{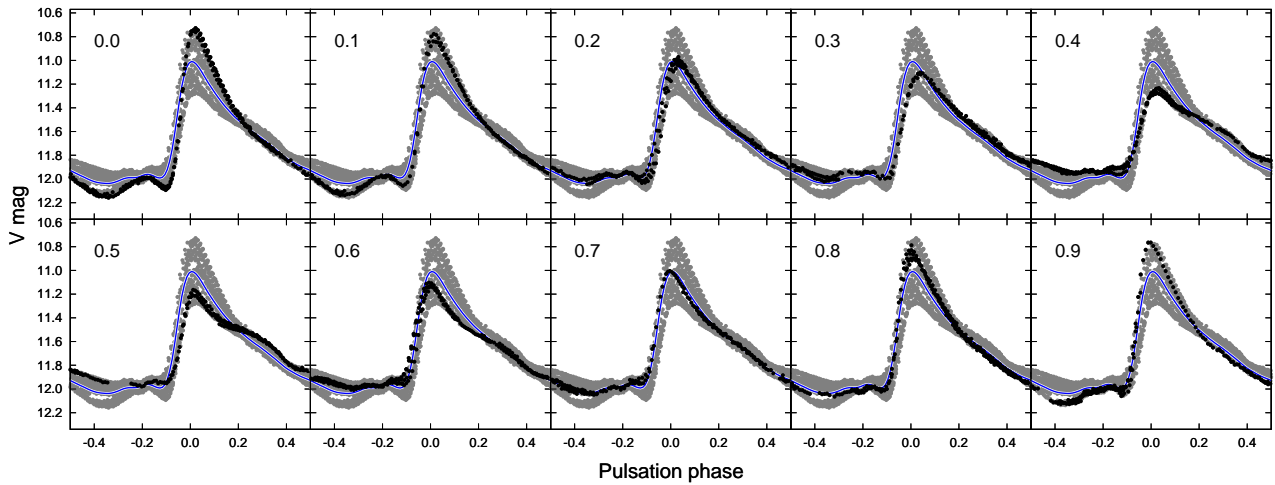}
\caption{Light-curve variation of RZ~Lyr during the Blazhko cycle.  The observations are plotted by gray dots; data belonging to a given Blazhko phase are highlighted in black; solid lines indicate the mean light-curve shape. The Blazhko phases ($\Psi$) are given in the top left-side corners in the panels.  }
\label{lc}
\end{figure*}

\begin{figure}
\begin{centering}
\includegraphics[width=7.5cm]{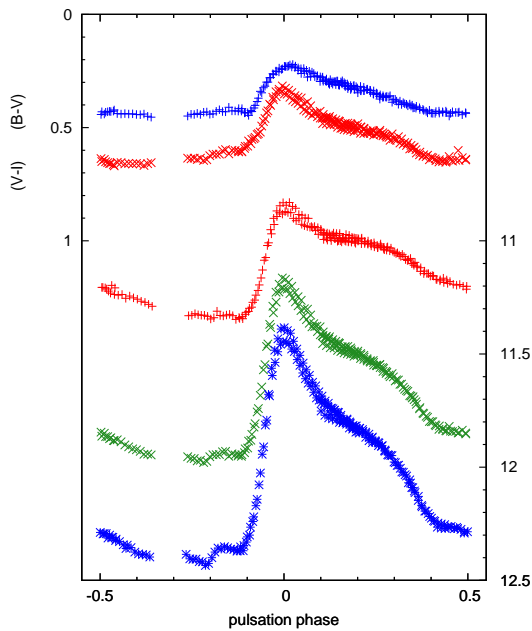}
\caption{Light and colour curves in the small-amplitude phase of the modulation ($\Psi=0.5$). A peculiar bump on the descending branch is pronounced in this Blazhko phase. These light curves look very much like the light variation of bump Cepheids (e.g.~$\eta$~Aql).}
\label{lc005}
\end{centering}
\end{figure}

\section{Light-curve variation during the Blazhko cycle}

The CCD $V$ light curves of RZ~Lyr in ten different phases of the modulation are shown in Fig.~\ref{lc}. The amplitude variation is as large as 0.8 mag and the phases of the maximum vary within about 0.08 pulsation phase  ($\sim$0.04 d).

In Blazhko stars showing large amplitude modulations, the pulsation light curve is hevily distorted  in the small-amplitude phase of the modulation, as in the extreme case of V445 Lyr \citep[][in preparation]{gu}, when the amplitude does not exceed 0.2 mag. In general, the amplitude of the light curve is significantly smaller in these Blazhko phases than the amplitude of any non-modulated Blazhko star with the same pulsation period and metallicity value.

The light curve of RZ~Lyr is also the most anomalous at the small-amplitude phase of the modulation. A pronounced bump appears on the descending branch centred around pulsation phase 0.25--0.30, when the pulsation amplitude is the smallest (see Fig.~\ref{lc005}). No similar feature of any other Blazhko or stable-light-curve RR Lyrae star has been ever observed, the detected bumps in RR Lyrae light curves do not appear earlier than phase 0.40 \citep[see e.g. Table 3 in][]{gc}. The double-peaked light curve, which shows a secondary maximum at the small-amplitude phase of the modulation of V455 Lyr \citep[][in preparation]{gu} may represent an extreme example of the bump phenomenon of RZ Lyr.
\begin{figure}
\includegraphics[width=8.6cm]{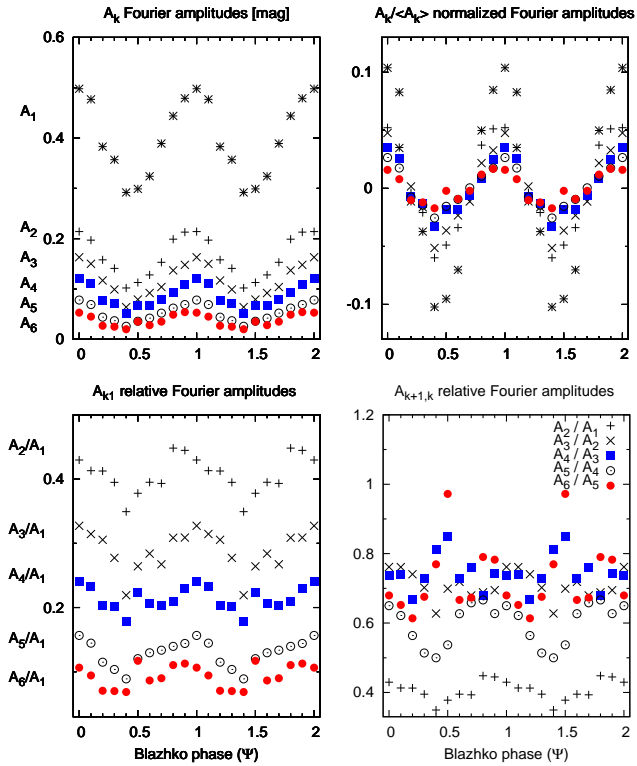}
\caption{Fourier amplitudes and amplitude ratios of the pulsation light curves in different Blazhko phases. The amplitudes of the harmonic components show parallel changes with decreasing amplitudes in increasing harmonic orders. The normalized amplitude variations (top right-hand panel) indicate that the absolute strength of the amplitude variation decreases also with increasing harmonic orders. The $A_k/A_1$ amplitude ratios of the fourth and sixth harmonic components (filled symbols) are anomalously large  at Blazhko phase 0.5 (bottom left-hand panel). This peculiar behaviour is even more evident if the  $A_{k+1}/A_{k}$ amplitude ratios  are plotted (bottom right-hand panel). }
\label{four}
\end{figure}

\begin{figure}
\begin{centering}
\includegraphics[width=8.6cm]{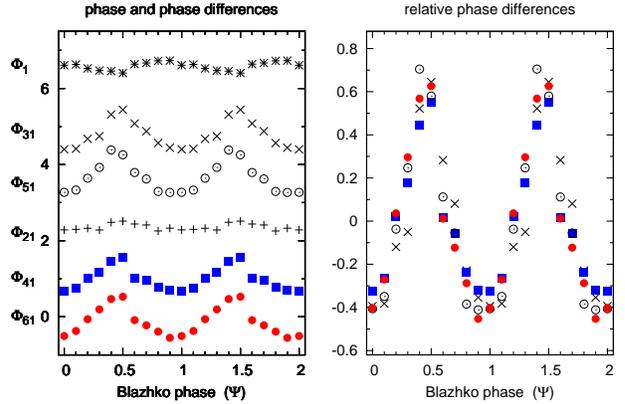}
\caption{Fourier phase of $f_0$ and phase differences ($\Phi_{k,1}$) of the higher-order harmonic components of the pulsation light curve in different phases of the modulation. The phase variation of the $f_0$ and $2f_0$ components ($\Phi_1, \Phi_{2,1}$) is only about 0.2 rad, while the phase variations of the higher harmonic-order components are as large as 1 rad. The amplitudes of the phase variations of the higher-order components are somewhat different as shown in the right-hand panel, which plots the absolute values of the phase difference variations. The amplitude of the variations in  $\Phi_{41}$ is the smallest.  }
\label{fourfd}
\end{centering}
\end{figure}

The bump on the descending branch on the small amplitude light curve of RZ Lyr is reminiscent to the light curves of bump Cepheids \citep{bc}. The periods of the fundamental and the second-overtone radial modes are in 3:1 resonance in these, double-mode,  single-periodic pulsators \citep{sm0}, resulting the bump in the light curve. The bump is also detected in the colour curves both in bump Cepheids and in RZ~Lyr, indicating that the temperature-decrease is less steep during the bump phase. Therefore, we speculate, whether a similar explanation for the appearence of the bump in RZ~Lyr also holds. According to linear and non-linear nonadiabatic models (G. Kov\'acs and Z. Koll\'ath private communications) the period of the fifth-overtone mode, which is, however, strongly damped in RR Lyrae stars, may be in 3:1 resonance with the fundamental-mode period in metal deficient variables if the luminosity is low enough ($L \approx 35-50 L_{\odot}$). If this were indeed the case, that is, a 3:1 resonance condition were temporally fulfilled in the low-amplitude phase of the modulation, then a peculiar behaviour of the amplitude and phase of the third-harmonic-order component of the pulsation frequency would be expected to be detected when the bump is pronounced. 

Figs.~\ref{four} and  \ref{fourfd} show the amplitude and phase variations of the six lowest-order pulsation components ($f_0, 2f_0,.. 6f_0$) in ten different phases of the modulation. An anomalous behaviour of the Fourier components at the low-amplitude phases of the modulation ($\Psi\sim 0.5$) is indeed detected, however, it is the most prominent in the amplitudes of the fourth and the sixth-order components (shown by filled symbols in the figures). Each amplitude ratio has a minimum value at $\Psi= 0.4$ but with different depths. The minima  of the variations of $A_2/A_1$, $A_3/A_1$ and  $A_5/A_1$ are deeper than the minima of  $A_4/A_1$ and $A_6/A_1$. Moreover, $A_4/A_1$ and $A_6/A_1$ have anomalously large values at  $\Psi= 0.5$ (bottom left-hand panel in Fig.~\ref{four}). Consequently, the relative amplitudes of the consecutive orders $A_4/A_3$ and $A_6/A_5$ have maxima at  $\Psi=$ 0.4--0.5,  while $A_2/A_1$, $A_3/A_2$ and $A_5/A_4$ show minima in this phase (bottom right-hand panel in Fig.~\ref{four}).

As for the phases of the harmonic components of the pulsation frequency, $\Phi_{21}$ hardly varies relative to $\Phi_{1}$, while the higher-order phase differences show $\sim$1 rad amplitude variations with a maximum at $\Psi=$ 0.4--0.5. The amplitudes  of the higher-order phase-difference variations differ slightly,  $\Phi_{41}$ exhibits the smallest-amplitude variation.

Consequently, if resonance causes the appearence of the bump, then it is most probably a 4:1 and not a 3:1 resonance. Recently, a  9:2 (4.5:1) half integer resonance with higher-order (8th--10th) radial modes have been suggested to explain the period doubling behaviour of some Blazhko stars observed by the $Kepler$ space telescope \citep{pd,pdd}. These higher-order modes have also been shown to be less damped as the third--seventh order modes. Their periods can match the 4:1 integer resonance condition in certain cases, as well. Therefore, we propose that a 4:1 resonance with a higher-order radial mode explains the bump-shape character of the light and colour curves of RZ~Lyr during the small-amplitude phase of the modulation. The feasibility of such a scenario has to be checked, however, by detailed hydrodynamical modelling.

Another, natural explanation for the bump would be a shock-wave propagation through the atmosphere, as shock waves are responsible for the appearence of the bumps and humps observable at around 0.7 and 0.9  pulsation phases \citep{gc,c00}, which are common features of both Blazhko and non-Blazhko RR Lyrae stars. However, no hydrodynamic study of the pulsation of RR Lyrae stars points to the existence of any shock wave at $\sim$0.3 pulsation phase, when the stellar radius is the largest.

\section{Frequency analysis of the CCD observations}

\begin{figure}
\begin{centering}
\includegraphics[width=8.8 cm]{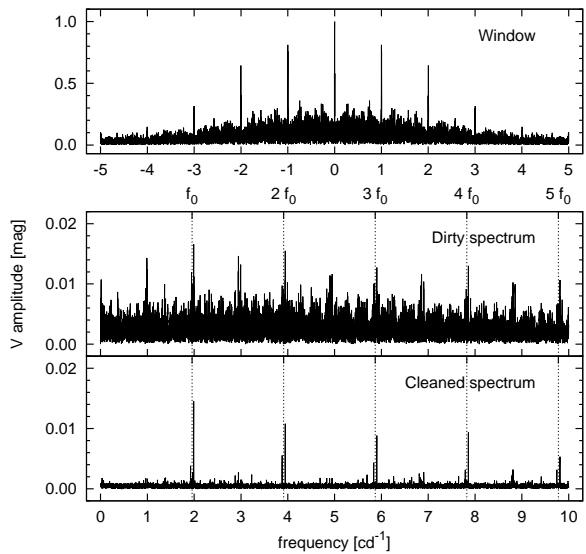}
\caption{The spectral window, the dirty and cleaned spectra of the data prewhitened for the pulsation and the main modulation components ($kf_0, kf_0\pm f_m$) obtained with the CLEAN Algorithm are shown in the top, middle and bottom panels, respectively. Although the daily alias components dominate the window structure, based on the cleaned spectrum the residuals can be identified as a secondary modulation, as the frequency components are located symmetrically at both sides of the main pulsation frequency components.}  
\label{sp}
\end{centering}
\end{figure}

\begin{figure}
\begin{centering}
\includegraphics[width=8.3 cm]{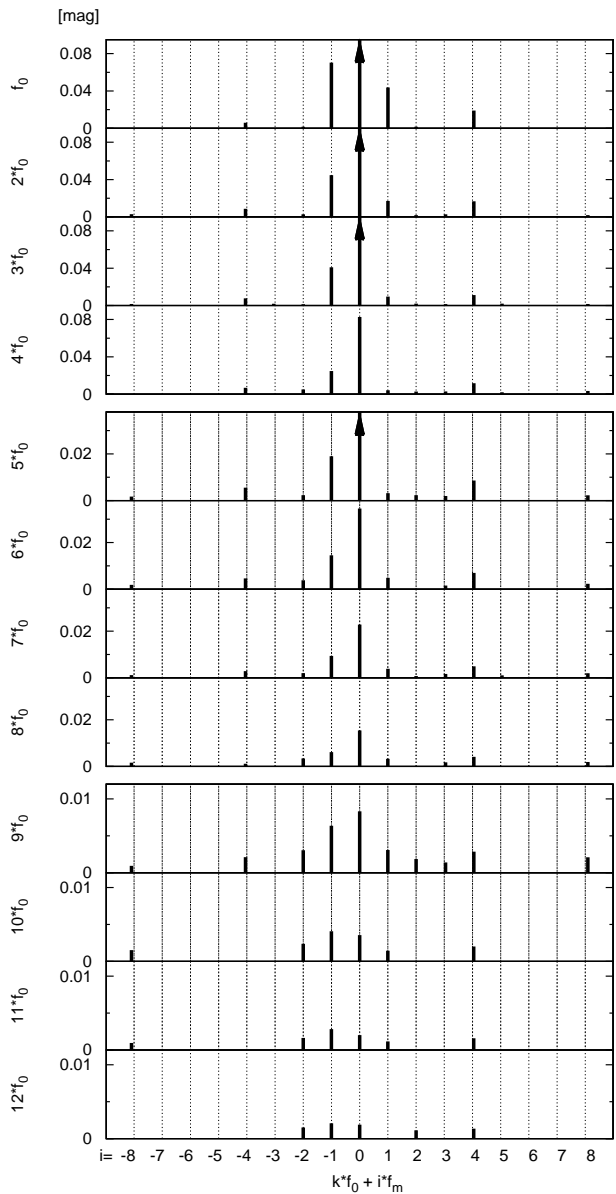}
\caption{Schematic plot of the CCD $V$ light-curve solution. The detected frequencies are indicated by vertical lines, with length corresponding to their amplitudes, in the vicinity of the pulsation frequency and its harmonics ($kf_0$). Note the different magnitude scales of the plots. The arrows indicate that the amplitudes of $f_0$,$2f_0$,$3f_0$ and $5f_0$ are out of the ranges. Several modulation components are present in each harmonic order, the most prominent features are the triplets with $f_\mathrm{m}=0.0082 $d$^{-1}$ (121 d) separation.}
\label{fr}
\end{centering}
\end{figure}

\begin{figure}
\begin{centering}
\includegraphics[width=8.8 cm]{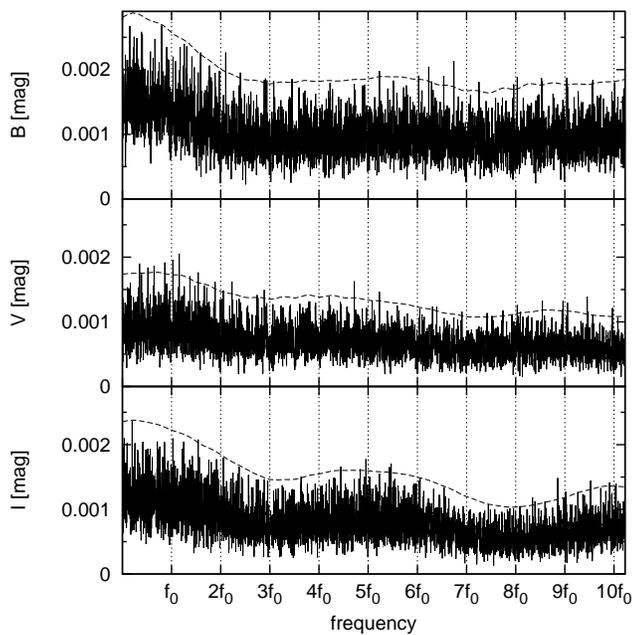}
\caption{Final residual spectra in the $B,V$ and $I_C$ bands and their smoothed,  3$\sigma$ levels defined as 3 times of the mean level of the residual spectrum are shown. There is no clear evidence of any additional frequency component.  }
\label{ressp}
\end{centering}
\end{figure}

The frequency analysis is performed using the MUFRAN package \citep{mufran}, the CLEAN Algorithm \citep{clean}, the GNUPLOT\footnote{http://www.gnuplot.info} utilities and a nonlinear fitting algorithm developed by \'A S\'odor. To a first approximation, the frequency spectrum of the data can be described by frequency triplets corresponding to the $\sim$121-d ($f_\mathrm{m}=0.0082$ d$^{-1}$) modulation of the pulsation light curve. 

A detailed analysis of the residuals reveals several additional frequency components. Fig.~\ref{sp} shows the spectral window and the dirty and cleaned Fourier amplitude spectra of the data prewhitened with the pulsation and the main modulation components (triplets). According to the cleaned spectrum, the residual can be described as a secondary modulation with components at around $\pm 4 f_\mathrm{m}$ separations from the main pulsation components.  22 frequencies matching this secondary modulation sequence have amplitudes larger than the 3.5 $\sigma$ level of the residual spectrum in at least one band; among these, the ratios of the observed amplitudes over their formal errors are larger than 4.0 for 21 components in at least two bands.

15 pulsation frequency components are detected, however, $14f_0$ and $15f_0$ are only marginally significant with amplitudes corresponding to 3.0, 2.0, 1.8 and 3.8, 1.9, 2.3 times the mean level of the residual spectra in the $B, V$ and $I_C$ bands, respectively. 25 and 22 modulation components belonging to the $\pm f_\mathrm{m}$ and $\pm 4f_\mathrm{m}$ sequences are identified; $f_\mathrm{m}$ and $4f_\mathrm{m}$ are also detected. A long-term drift ($f'\approx0.002$ d$^{-1}$) remains in the residuals after prewhitening the data with the above frequencies, which has, most probably, instrumental instead of stellar origin.

After these prewhitening steps, the residuals still show the largest-amplitude signals in the vicinity of the pulsation frequencies. Signals appear in these spectra at around $kf_0\pm 2f_\mathrm{m}$, $\pm 3f_\mathrm{m}$ and $+5f_\mathrm{m}$; the amplitudes of 26 components belonging to these series are larger than the 3.5$\sigma$ level of the residual spectrum in at least two bands, while in 6 less certain cases the amplitudes are larger than the 3.0$\sigma$ level in one band.

Finally, 11 components at around $\pm 8f_\mathrm{m}$ separation from the pulsation components are detected with 3.5$\sigma$ level of the residual spectrum in at least two bands, while in 5 less certain cases the amplitudes are larger than the 3.0$\sigma$ level in one band.

Fig.~\ref{fr} delineates the result of the light-curve analysis schematically. The detected frequencies around the pulsation components are drawn by their amplitudes obtained for the $V$ light-curve solution. The amplitude distribution of the modulation components shows a unique feature. The amplitudes of the frequencies at $\pm 4 f_\mathrm{m}$ separation are significantly larger than the amplitudes of the  $\pm 2 f_\mathrm{m}$ components, moreover, the frequencies at $\pm 8 f_\mathrm{m}$ separations have similar amplitudes as the $\pm 2 f_\mathrm{m}$ quintuplet components. This is in high contrast with the results of all the detailed analyses of Blazhko light curves where quintuplets and/or higher order multiplet components appear \citep{hu08,ju08,dmc,po10,ch10,gu11,ko11}. The amplitudes of the multiplet components typically decrease with increasing multiplet orders  according to the analysis of all the well observed Blazhko stars' light curve. Another general feature of the modulation is that, if higher multiplet-order components are detected, then the entire `spectrum' of the multiplet is more or less equally populated. These properties are  valid especially
in the lower pulsation orders. 

The discrepant behaviour of the amplitudes of the multiplets observed in RZ~Lyr suggests that the $\pm 4 f_\mathrm{m}$ and $\pm 8 f_\mathrm{m}$ components belong to an independent modulation quintuplet. The components of this secondary modulation are close to the 4th and 8th order multiplets of $f_\mathrm{m}$ but do not exactly equal with these frequencies. The residual scatter of the light-curve solution corresponding to this scenario is smaller by $\sim$10 per cent than the rms of the one-modulation solution as documented in Table~\ref{ff}. Adopting two independent modulation components ($f_\mathrm{m1}$ and $f_\mathrm{m2}$), the frequency solution of the $V$ data yields a frequency value for the secondary modulation that differs by more than 30$\sigma$ from the exact $\pm 4 f_\mathrm{m}$ value. 

It is also important to note that the amplitudes of the $kf_0 - f_\mathrm{m1}$ negative components of the main triplets are larger than the amplitudes of the $kf_0 + f_\mathrm{m1}$ positive components, while the secondary modulation shows an opposite character ($kf_0 + f_\mathrm{m2}>kf_0 -f_\mathrm{m2}$). The dissimilarity of the $f_\mathrm{m1}$ and the $\sim 4f_\mathrm{m1}=f_\mathrm{m2}$   modulations supports interpreting them as independent modulations.

\begin{table}
\caption{Independent frequencies of the CCD light-curve solution}
 \label{ff}
  \begin{tabular}{@{\hspace{5pt}}c@{\hspace{5pt}}c@{\hspace{5pt}}cr@{\hspace{5pt}}r@{\hspace{5pt}}r}
  \hline
$f_0$&$f_\mathrm{m1}$&$f_\mathrm{m2}$&\multicolumn{3}{c}{Residual scatter [mag]}\\
     &        &        & \multicolumn{1}{c}{$B$}&\multicolumn{1}{c}{$V$}&\multicolumn{1}{c}{$I_\mathrm{C}$}\\
\hline
1.9560717(4)&  0.008305(2)   &  4$fm_1$  & 0.0140& 0.0103& 0.0110 \\
1.9560667(4)&  0.008243(3)   & 0.033330(7)\,\,& 0.0121& 0.0090& 0.0102\\
&\multicolumn{2}{r}{$4f_\mathrm{m1}=$0.032972(12)}&&&\\
\hline
\multicolumn{6}{l}{\footnotesize{errors in parentheses indicate formal 1$\sigma$ uncertainties}}\\
\end{tabular}
\end{table}

\begin{table*}
\caption{Frequencies, amplitudes (in mag) and phases (in rad) of the $B$, $V$ and $I_\mathrm{C}$ light-curve solutions of RZ~Lyr. The $A_0$ values are the magnitude zero points relative to comparison star `a'. The full table is available as Supporting Information with the online version of this paper.}
 \label{freq}
  \begin{tabular}{ccllllll}
\hline
               &               &\multicolumn{2}{c}{$B$} &\multicolumn{2}{c}{$V$} &\multicolumn{2}{c}{$I_\mathrm{C}$}      \\
Identification &$f$ [cd$^{-1}$]&Amplitude     &Phase    &Amplitude    &Phase     &Amplitude    &Phase\\
\hline
$A_0$          & &0.548&&0.658&&0.696&\\
\\
$1f_\mathrm{0}$         &1.956067       &0.5110(4)     &1.521(1) &0.3931(3)    &1.481(1)  &0.2395(3)    &1.328(1)\\
$2f_\mathrm{0}$         &3.912133       &0.2000(4)     &5.305(2) &0.1582(3)    &5.313(2)  &0.0982(4)    &5.273(4)\\
$3f_\mathrm{0}$         &5.868200       &0.1382(4)     &2.884(3) &0.1110(3)    &2.885(3)  &0.0720(4)    &2.879(5)\\
...\\
$f_\mathrm{m1}$       &0.008243       &0.0042(4)     &1.9(1)   &0.0026(3)    &1.5(1)    &0.0012(4)    &0.5(3)\\
$2f_\mathrm{m1}$      &0.016486       &0.0031(4)     &3.3(1)   &0.0026(3)    &3.6(1)    &0.0021(4)    &3.7(2)\\
$f_\mathrm{m2}$       &0.033330       &0.0045(4)     &4.5(1)   &0.0038(3)    &4.64(9)   &0.0042(4)    &4.6(1)\\
$1f_\mathrm{0}-f_\mathrm{m1}$  &1.947824       &0.0903(4)     &3.496(4) &0.0704(3)    &3.511(4)  &0.0452(3)    &3.547(8)\\
$2f_\mathrm{0}-f_\mathrm{m1}$  &3.903890       &0.0554(4)     &0.978(8) &0.0448(3)    &0.992(7)  &0.0284(4)    &1.02(1)\\
...\\
$1f_\mathrm{0}+f_\mathrm{m1}$  &1.964310       &0.0570(4)     &0.273(7) &0.0437(3)    &0.295(7)  &0.0275(3)    &0.30(1)\\
$2f_\mathrm{0}+f_\mathrm{m1}$  &3.920376       &0.0209(4)     &4.34(2)  &0.0171(3)    &4.36(2)   &0.0105(4)    &4.34(3)\\
...\\
$1f_\mathrm{0}-f_\mathrm{m2}$  &1.922737       &0.0097(4)     &2.93(4)  &0.0059(3)    &2.92(6)   &0.0018(4)    &3.0(2)\\
$2f_\mathrm{0}-f_\mathrm{m2}$  &3.878803       &0.0121(5)     &0.41(4)  &0.0087(3)    &0.47(4)   &0.0053(4)    &0.47(7)\\
...\\
$1f_\mathrm{0}+f_\mathrm{m2}$  &1.989397       &0.0255(5)     &4.00(2)  &0.0190(3)    &4.06(2)   &0.0118(4)    &4.16(3)\\
$2f_\mathrm{0}+f_\mathrm{m2}$  &3.945463       &0.0202(4)     &1.39(2)  &0.0168(3)    &1.42(2)   &0.0111(4)    &1.55(4)\\
...\\
\hline
\end{tabular}
\end{table*}

\begin{figure}
\begin{centering}
\includegraphics[width=7. cm]{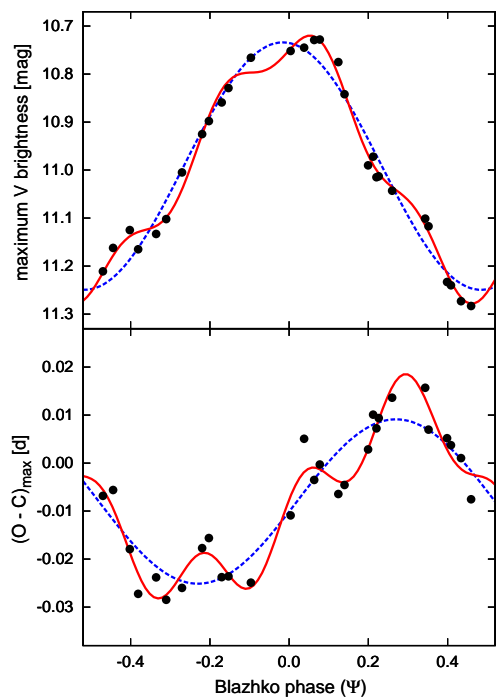}
\caption{Maximum-brightness and maximum-phase variations according to the CCD $V$ maximum observations given in Table~\ref{max}. First and fourth-order fits to the data are also drawn. The fourth-order components of the fits have well pronounced amplitudes indicating that, beside the 121-d primary modulation, a $\sim$30-d periodicity in the light-curve variation is also present.}
\label{max}
\end{centering}
\end{figure}

The maximum-brightness and the maximum-phase values of the CCD data (Table~\ref{maxima}) are phased with the 121-d modulation period in Fig.~\ref{max}. First and  fourth-order harmonic fits to the data are drawn in the Figures. Unambiguously,  the amplitudes of the fourth-order components of both the maximum-brightness and maximum-phase data are significantly larger than the amplitudes of the second and third harmonic-order components. This result confirms that the modulation  is not singly periodic, and that the period of the secondary modulation is very close to $P_\mathrm{m}/4$.

We thus conclude that RZ~Lyr, similarly to CZ Lac \citep{czl}, exhibits two modulations with modulation frequency values being very close to the 1:4 ratio. Table~\ref{freq} gives the frequencies of the light-curve solution and the amplitudes and phases of the frequencies for the $B,V$ and $I_\mathrm{C}$ time series. The complete table is available from the electronic version of this paper. The light variation of RZ Lyr is described by three independent frequencies $f_0, f_\mathrm{m1}$ and $f_\mathrm{m2}$. In this representation, the modulation frequencies at around $\pm 3f_\mathrm{m}$ and $\pm 5f_\mathrm{m}$ are fitted as combinations of the two modulations, i.e.  $k(-f_\mathrm{m1}+f_\mathrm{m2})$ and $k(f_\mathrm{m1}+f_\mathrm{m2})$.

The rms scatters of the full light-curve solutions that indicate both the goodness of the fits and the accuracy of the observations are 0.012, 0.009 and 0.010 mag for  the $ B, V$, and $I_C$ data, respectively. The residual spectra and their  3$\sigma$ levels, which are calculated as 3 times the smoothed mean values of the residual spectra, are shown in Fig.~\ref{ressp} for the three bands.

\section{Physical parameters and their variations during the modulation}

\begin{table*}
\centering
\caption{The mean physical parameters and their uncertainties are derived from the $BVI_\mathrm{C}$ mean pulsation light curves of RZ~Lyr by the inverse photometric method (IPM) using static atmosphere models with \hbox{[Fe/H] =$-1.85$} metallicity. The luminosity is selected to match the evolutionary possible values for a metal-deficient horizontal-branch star and the mean physical parameters are required to fit the pulsation period via the pulsation equation. }
\label{tbl:ipm_totalmean}
\begin{tabular}{cccccccc}
\hline
$\mathfrak{M}$       & $d$         & $M_V$           & $R$           & $L$          & $T_\mathrm{eff}$ \\ 
$\mathfrak{M}_\odot$ & (pc)        & (mag)           & ($R_\odot$)   & ($L_\odot$)  &        (K)       \\
\hline
$0.66\pm0.04$        & $1380\pm60$ & $0.739\pm0.004$ & $5.12\pm0.20$ & $47.5\pm2.5$ & $6670\pm150$     \\
\hline
\end{tabular}
\end{table*}

The variations of the pulsation-averaged atmospheric parameters of RZ~Lyr during the Blazhko cycle are derived using the inverse photometric Baade--Wesselink method \citep[IPM;][]{ip}. These parameters are the effective temperature ($T_\mathrm{eff}$), luminosity ($L$), radius ($R$) and effective surface gravity ($\log g$). The IPM needs only multicolour photometric time series and synthetic colours from static atmosphere models \citep{kurucz} as input. We applied this method already for five modulated RRab stars (MW~Lyr -- \citealt{mw2}; DM~Cyg -- \citealt{dmc}; RR~Gem, SS~Cnc -- \citealt{sodor_aipc}; CZ~Lac -- \citealt{czl}) succesfully.

\subsection{Constant parameters}
\label{stpar}

In the first step, the  Blazhko-phase independent parameters of RZ~Lyr have to be determined. These are the metallicity ([Fe/H]), mass ($\mathfrak{M}$), distance ($d$) and the interstellar reddening [$E(B-V)$]. 

The metallicity of RZ~Lyr was measured by \cite{su} ($\Delta$S $=9.89$) and \cite{la} ([Fe/H]$=-2.13$).  On a combined, homogenized metallicity scale \citep{ju95,jk96}, these observations correspond to $-1.84$ and $-1.85$ [Fe/H] values, respectively. Thus, we accept [Fe/H]$=-1.85$ for the metallicity of RZ~Lyr.

As it was shown in \cite{bl}, the well-defined  mean pulsation light curves of Blazhko stars yield correct metallicity value according to the [Fe/H]$(P,\Phi_{31})$ formula \citep{jk96}. This is not the case, however, for RZ~Lyr. The photometric metallicity defined by the mean $V$ light curve is $-1.44$. The anomalous light-curve shape at the small-amplitude phase of the modulation (see details in Section~\ref{lc}) may account for the incorrect photometric metallicity estimate. If data belonging to Blazhko phases 0.4--0.6 (the light curves in these phases show an anomalous bump on the descending branch) are removed from the data set, then the photometric metallicity is $-1.65$. This is closer to the spectroscopic metallicity value, but it is still 0.2 dex more metal-rich than expected. We thus conclude that the applicability of the metallicity formula for the light curves of Blazhko stars is questionable in some cases, even if the mean light curve is well-defined. 

The interstellar reddening towards the direction of RZ~Lyr  is $0.09$~mag according to the extinction maps of \cite{schlegel}.  \cite{blanco} derived   $E(B-V)=0.07$~mag for  RZ~Lyr from the near-minimum-light $B-V$ colour.  The average of these two estimates, $0.08$~mag, is accepted for the reddening of RZ~Lyr. The unreddened colours of RZ~Lyr, which also depend on the  uncertainties of the magnitudes of the comparison star (Table~\ref{comp}), are supposed to be accurate within about 0.03 mag.

The next two constant parameters we have to determine are the  mass ($\mathfrak{M}$) and the distance ($d$). In principle, IPM is capable to drive these parameters by the application of the method on the mean pulsation light curve. 

However, if the fitting process is allowed to adjust these parameters unconstrained, then the resulting mass and luminosity values are too high if compared to  horizontal-branch (HB) evolutionary model results. Maybe, as a result of the bump-shaped light curve at the minimum-amplitude phase of the modulation, the mean light curve of RZ Lyr remains also somewhat anomalous. Therefore, the luminosity is constrained to match the oxygen-enhanced HB evolutionary models of \cite{dorman92}. The possible luminosity range of the instability strip of the [Fe/H]=$-1.78$ HB models is \hbox{45--57$\,L_\odot$} \cite[][fig.~4]{dorman92}.
A modified pulsation equation \citep[][eq.~2]{ip} is applied to link the mass, radius, temperature and period values during the fitting process of the IPM. These constraints yield \hbox{$\mathfrak{M}=0.62$\,--\,$0.83\,\mathfrak{M}_\odot$} for the possible mass range.  The HB evolutionary models contradict, however, the solutions with masses above $0.7\,\mathfrak{M}_\odot$, as these model tracks do not enter the instability strip at the given metallicity. Therefore, \hbox{$L=47.5 \pm 2.5\,L_\odot$} and \hbox{$\mathfrak{M}=0.66\pm0.04\,\mathfrak{M}_\odot$}  are accepted for the possible luminosity and mass ranges of RZ~Lyr. These parameters  result in a \hbox{$d=1380$\,pc} distance estimate with \hbox{$\pm60$\,pc} uncertainty depending on the  accepted values/uncertainties of the luminosity, the $V$ light curve magnitude zero point and the interstellar absorption [$A_\mathrm{V}=3.14E(B-V)$].

Applying these constraints, the goodness of the IPM fits are only marginally worse  than in the let-free case.

The pulsation- and modulation-phase-averaged mean radius, temperature and absolute visual brightness ($M_\mathrm{V}$) of RZ~Lyr are derived applying the IPM for the $BVI_\mathrm{C}$ mean pulsation light curves with fixed luminosity, mass and distance values corresponding to the above discussed constraints.

Similarly to the previous analyses, the IPM code is run with four different internal settings \citep[for details see][table~1]{ip} to estimate the inherent uncertainty of the method.

The results for the mean physical parameters of RZ~Lyr are summarized in Table~\ref{tbl:ipm_totalmean}. The error ranges are estimated by taking into account the inherent error of the method, the uncertainties of the input parameters and the possible ranges of the constrained parameters.

\subsection{Blazhko-phase dependent parameters}

\begin{figure}
 \begin{centering}
  \includegraphics[width=90mm]{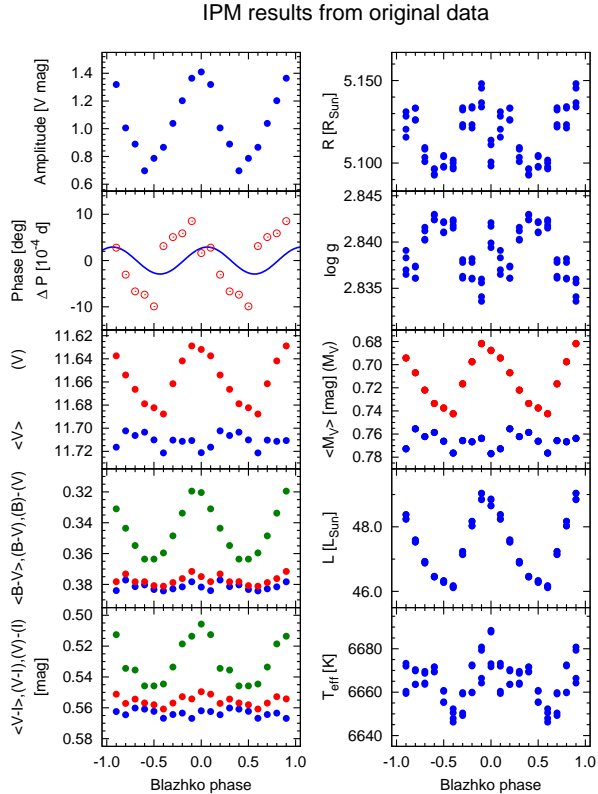}
\caption{Variations of the mean observed (left-hand panels) and the derived physical parameters (right-hand panels) of RZ~Lyr in different phases of the primary ($\sim$121 d) modulation. The secondary ($\sim$30 d) modulation has not been removed from the data, this modulation accounts partially for the scatter e.g. in the magnitude averaged V brightness and the radius and log\,$g$ variations. Magnitude- and intensity-averaged brightnesses and colours are denoted by angle and round brackets, respectively. From top to bottom, the left-hand panels show: amplitude modulation -- total pulsation amplitude; pulsation-phase or period modulation -- variation of the phase of the $f_\mathrm{p}$ pulsation component (empty circles) and deviation of the instantaneous pulsation period from the mean pulsation period (continuous line); pulsation-averaged $V$ magnitude; pulsation-averaged $B-V$ colour; pulsation-averaged $V-I_\mathrm{C}$ colour. From top to bottom, the right-hand panels show the pulsation averages of the following physical parameters: radius; surface gravity; absolute visual magnitude; luminosity; effective surface temperature. \label{fig:ipresorig}}
 \end{centering}
\end{figure}
Fixing the Blazhko-phase independent physical parameters to their values determined in Section~\ref{stpar}, the IPM is run using the light curves of RZ~Lyr in 10 different phases of the modulation. In this way, the  radius, luminosity, temperature and surface-gravity changes during the pulsation in different phases of the modulation
are determined. The pulsation-averaged mean values of the variations of these parameters characterize the global mean changes in the stellar parameters during the modulation cycle. Similarly to the IPM results of the previous investigations of Blazhko stars, the actual values of the fixed or constrained physical parameters influence only the averages of the other physical parameters, and  affect their variations during the Blazhko cycle marginally. A small amplitude difference between the solutions is detected if the input parameters are varied within their possible ranges (Table~\ref{tbl:ipm_totalmean}), but the relative sign of their variations is unaffected.

As RZ~Lyr has two independent modulation periods, and the shorter one is approximately, but not exactly, one quarter of the longer one, the  mean physical parameter variations are investigated both on the original data supposing a single-periodic modulation, and on synthetic data, which separate the primary and secondary modulation components.
The separation of the modulations is carried out by generating synthetic light curves from the light-curve solution given in Table~\ref{freq} using pulsation frequencies and modulation terms  belonging to only one of the modulations, namely $\pm f_\mathrm{m1}, \pm 2f_\mathrm{m1}$ and $\pm 3f_\mathrm{m1}$ in one case, and $\pm f_\mathrm{m2}$ and $\pm 2f_\mathrm{m2}$ in the other case. The minute coupling modulation terms are omitted, as they are not periodic with any of the modulations. Considering their rather small amplitudes, the omission of these terms has  a negligible effect on the results (a similar treatment was applied in the IPM analysis of CZ~Lac \citep{czl}, another multiple-periodic Blazhko star).

The IPM results of the original time series are plotted against the phases of the 121.4-d dominant modulation in Fig.~\ref{fig:ipresorig}. As a result of the secondary modulation, which has a period approximately one quarter of the primary modulation period, the resulting curves have, in principle, wavy shapes but the 10 phase bins of the main modulation are insufficient to resolve these features. Therefore, most of the  phase-bin-to-phase-bin variations (these are the most prominent in the magnitude-averaged brightness, radius, and log\,$g$ variations) are probably real features connected to the secondary modulation and  not the consequences of the inherent noise of the method and/or the inaccuracy of the input light curves.

In Fig.~\ref{fig:ipresorig},  the observed (left-hand panels) and derived (right-hand panels) variations of the pulsation-period-averaged parameters during the Blazhko cycle of the main modulation are plotted. The observed parameters (from top to bottom) are the full amplitude, the $\Phi_{1}$  phase variation of the main pulsation frequency ($f_0$), and the different averages of the light and colour curves according to magnitude and intensity-scale representations. The solid line drawn in the phase-variation panel indicates the pulsation-period variation corresponding to the observed phase variation of $f_0$. The total range of the period variation is about 0.0006~d during the Blazhko cycle. No significant phase lag between the amplitude and period variation of the pulsation light curve is detected during the main, 121-d modulation.

\begin{figure*}
 \begin{centering}
\includegraphics[width=90mm]{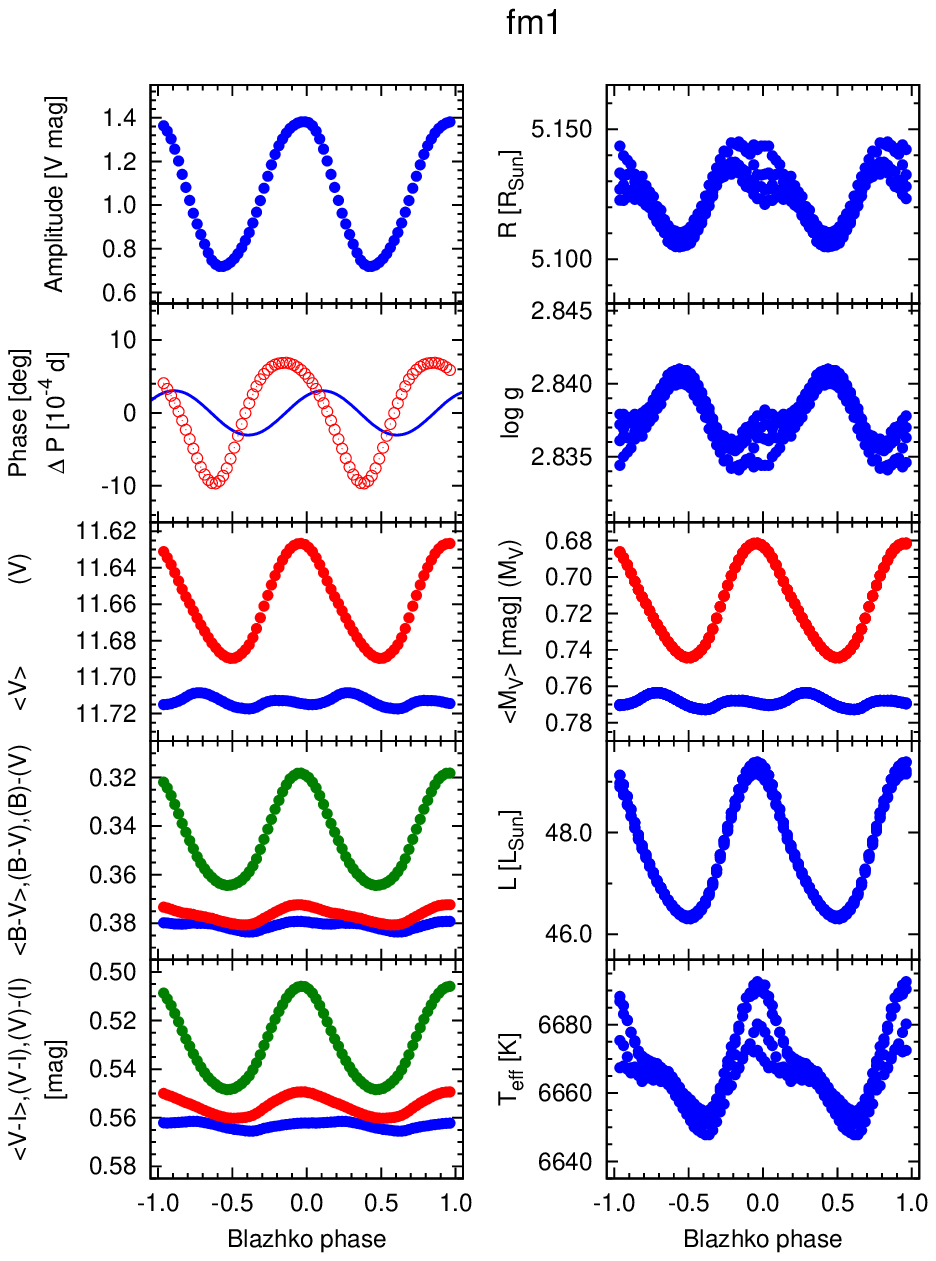}\includegraphics[width=90mm]{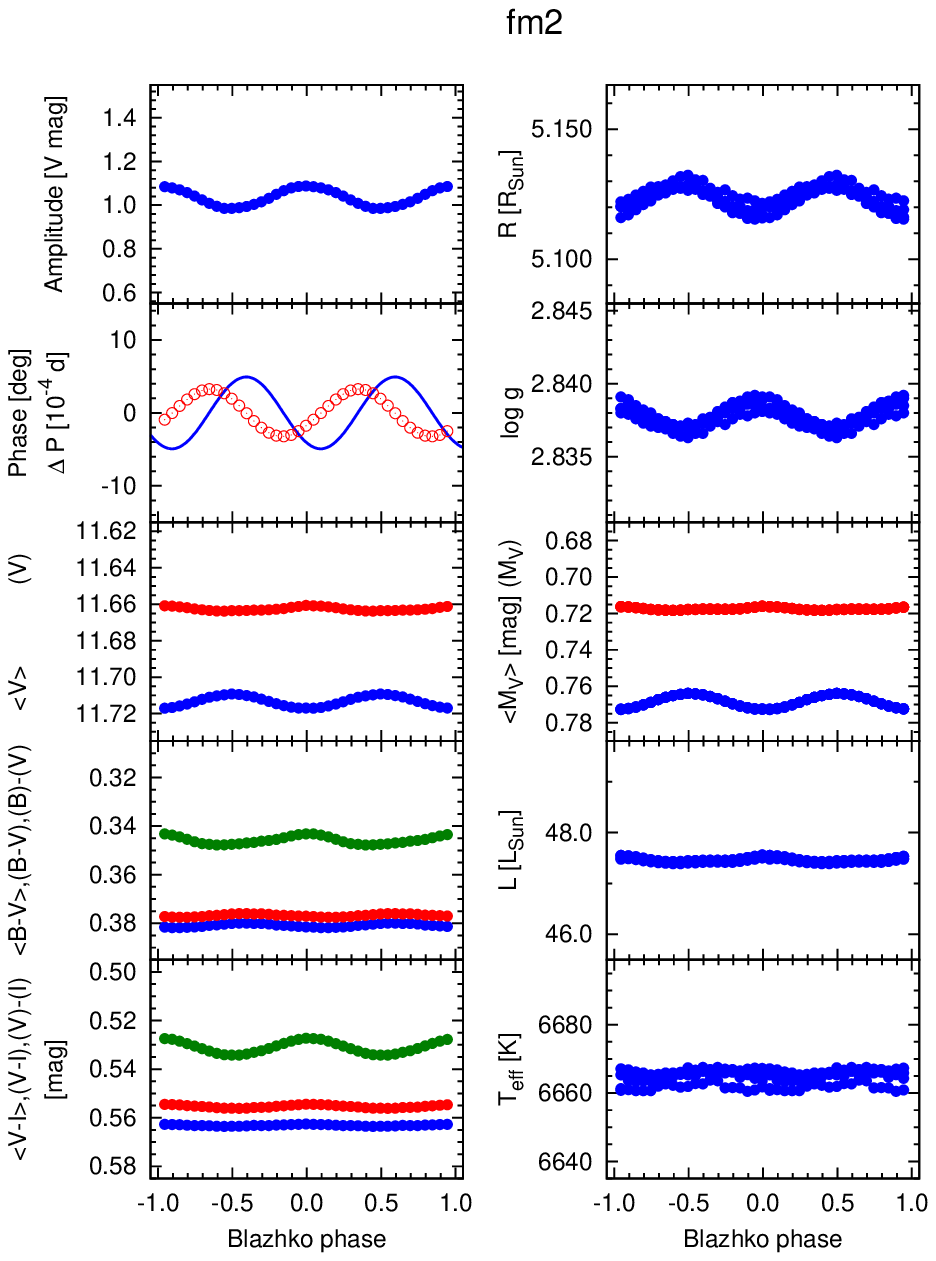}\\
\caption{Variations of the physical parameters of RZ~Lyr according to the phases of the primary ($\sim$121 d) and secondary ($\sim$30 d) modulations. The two modulations are separated using the appropriate frequencies of the light-curve solution given in Table~\ref{freq}. The panels are the same as described in the caption of Fig.~\ref{fig:ipresorig}. }
\label{fig:ipressynth}
\end{centering}
\end{figure*}

The mean physical-parameter variations shown in the right-hand panels of Fig.~\ref{fig:ipresorig} indicate an extremely large (6.3 per cent) luminosity variation and modest, 0.8 and 0.4 per cent radius and temperature changes, respectively.
The variations in the mean radius, luminosity and temperature are all in line with the amplitude variation of RZ~Lyr, that is, these parameters have maximum values when the pulsation amplitude is the largest.

The IPM solutions of the synthetic data show the variations of the physical parameters along the two modulation components separately (Fig.~\ref{fig:ipressynth}).
The results for the main modulation component agree well with the results obtained for the original light-curve data (Fig.~\ref{fig:ipresorig}), while the physical-parameter variations along the secondary modulation show a different behaviour. The amplitude variation is the opposite of the phase variation in this modulation, that is, when the amplitude is the largest the period is the shortest. No detectable temperature variation  and very small (0.2 per cent) luminosity change during the $\sim$30-d modulation are derived. The 0.01\,$R_\odot$  radius variation, which is in anti-phase with the amplitude variation and is in phase with the period variation is the most prominent change connected to this modulation component. It is also worth to note that amplitude variation dominates the modulation of the primary and phase modulation of the secondary modulations. The range of the period variation in the secondary modulation is larger compared to the range of the period variation during the primary modulation.

\section{Summary}
Although the number of Blazhko stars with observations allowing detailed analyses is increasing, we can still learn a lot about  the modulation phenomenon from these data. The analysis of RZ~Lyr disclosed some novelties of the Blazhko modulation, too. 
\begin{itemize}
\item Based on multi-epoch observations, a strict anticorrelation between the pulsation and modulation period changes is found with ${\mathrm d}P_0/{\mathrm d}P_{\mathrm m}=-2.4\times 10^{-6}$ or $({\mathrm d}P_0/P_0)/({\mathrm d}P_{\mathrm m}/P_{\mathrm m})=-5.6\times 10^{-4}$ ratio.
\end{itemize}
There is a tendency that the pulsation and modulation periods of Blazhko stars exhibit preferably opposite-direction changes, however, no strictly monotonic and stable period-change behaviour has been detected previously. The strong anticorrelation between the periods of RZ~Lyr suggests that the modulation period  cannot have an ad hoc value, its length and its variation is defined by the stellar parameters and their variations, as in the case of the pulsation period value.
\begin{itemize}
\item A pronounced bump on the descending branch of the light curve in the small-amplitude phase of the modulation is detected. Based on the similarity of the appearence of the bump to the light-curve shapes of bump-Cepheids we speculate that some resonance mechanism is responsible for the bump phenomenon in this case as well. Without a detailed hydrodynamical modelling, which is, however, out of the scope of the present paper, this possibility can neither be proved nor rejected. 
\end{itemize}
No RR Lyrae star with stable light curve is known that shows either period doubling \citep{pd} or a bump on the descending branch of the light curve at around 0.3 pulsation phase. Supposing that 4.5:1 and 4:1 half-integer and integer resonances are responsible for the intermittent occurrence of these phenomena in Blazhko stars, it can be concluded that the resonance conditions  never come true in a normal RRab star. These are unique, temporal specialities of some Blazhko stars at certain phases of the modulation. If this is indeed the case, then it means that the structure of a Blazhko star may  differ significantly  (at least in some Blazhko phases) from that of a normal RR Lyr star.
The detected changes of the mean physical parameters during the Blazhko cycle (as the  6 per cent change in the mean luminosity) may support such a scenario. 

The onset of period doubling is not strictly connected to any Blazhko phase in RR Lyr but the bump feature appears only in the small-amplitude phase of the modulation in RZ~Lyr. The modulation period of RR Lyr is $\sim40$ d, significantly shorter than the 121-d Blazhko period of RZ~Lyr. Moreover, the modulation itself is highly variable in RR~Lyr while it is rather regular in RZ~Lyr. Therefore, it is a plausible conclusion that while the long period and the regular behaviour of the modulation of RZ~Lyr favour the occurrence of modulation-phase locked realization of the resonance condition,
in RR~Lyr, which has a shorter Blazhko period and its consecutive modulation cycles are not fully repetitive, the  connection between the appearence of period doubling and modulation phase is only partial if there exists any.
\begin{itemize}
\item The modulation of RZ~Lyr is intrinsically doubly periodic. The secondary modulation period is close to but does not exactly equal with one quarter of the period of the 121-d main modulation.
\end{itemize}
Many Blazhko stars exhibit complex, not strictly periodic light-curve modulation. The modulations of these variables are most probably multi-periodic. The only detailed study of a double-modulation-period Blazhko star, CZ Lac, showed that there is a tendency that the periods/frequencies of the independent modulation components are close to being in resonance \citep{czl}. This is also the case in RZ~Lyr, where $f_\mathrm{m1}/f_\mathrm{m2}\approx 1:4$. It is also worth to note that the two modulations have different characters both in RZ Lyr and in CZ Lac. The explanation of the multi-periodic modulations is one of the greatest challenge for the theoretical elucidation of the Blazhko effect. 
\begin{itemize}
\item By the aid of the IP method, the changes in the mean values of the physical parameters during the modulation are determined. An extremely large, $\sim$3$\,L_\odot$ luminosity variation is detected; the star is the brightest at the large-amplitude phase of the modulation. The variations of the physical parameters connected to the primary and the secondary modulations show different behaviours. The pulsation amplitude,  period, stellar radius, luminosity and temperature all vary in phase with each other during the primary modulation. On the contrary, during the secondary modulation, the pulsation amplitude and period variations have opposite directions, and the radius variation follows the period, while the  marginal luminosity variation follows the amplitude changes.
\end{itemize}
According to the idea of \cite{st1,st2} a convective dynamo cycle is responsible for the light-curve modulation of Blazhko stars. The model gives predictions for the strengths of the period variation during the modulation as well as  for the phase relation of the period and radius variations. No phase (period) variation of variables with pulsation period around 0.5~d, and anticorrelated and correlated changes in the pulsation period and the radius variations are predicted to occur in hotter and cooler variables, respectively. The observed physical variations of MW Lyr \citep{mw2} and DM Cyg \citep{dmc} agree well with these predictions, while the complex modulation behaviour of  CZ Lac \citep{czl} fit in only partially with this scenario \citep{st3}.

The 0.51-d pulsation period of RZ~Lyr is close to the half-day critical period value, and indeed, its amplitude variation is much larger than the phase variation during the primary 121-d modulation. It is, however, not true for the secondary modulation, which is dominated by phase variation. Note that there are Blazhko stars showing negligible phase variations in the 0.55--0.60~d pulsation period regime, too (Konkoly Blazhko Survey II, Jurcsik et al. in preparation).

The phase relations between the period and radius changes are correlated with (dP/P)/(dR/R) ratios of  0.0011/0.006 and 0.0018/0.002  for the primary and secondary modulations, respectively. RZ~Lyr is found to be indeed hotter than MW~Lyr, which shows anticorrelated period and radius changes \citep{mw2}, but its temperature is the same or even somewhat warmer than the temperature of DM~Cyg \citep{dmc}, a smaller amplitude `twin' of MW~Lyr.

These results confirm only in part Stother's model, which has been seriously criticized by \cite{sm}. The large variety of the physical parameters of RR Lyrae stars makes it, however,  unlikely that any model could give universally valid, unique predictions that hold for the whole  possible metallicity and luminosity ranges of the variables. Mapping the physical parameter variations of a larger sample of Blazhko stars during the modulation cycle is required to reveal any connection between the amplitude and phase relations of these changes and the physical parameters of the stars.

\section*{Acknowledgments}
We thank G. Kov\'acs and Z. Koll\'ath for providing us with the 3:1 resonance solutions of linear and non-linear RR Lyr models. The financial support of OTKA grants K-68626 and K-81373 is acknowledged. We thank the many amateur and professional astronomers for participating in the efforts of compiling the GEOS data base, which provided a very usefull collection of data for this study. This research has made use of the SIMBAD database and ALADIN operated at CDS, Strasbourg, France.

\vfill
\newpage

\section{Appendix}

\begin{figure}
\begin{centering}
\includegraphics[width=8.6cm]{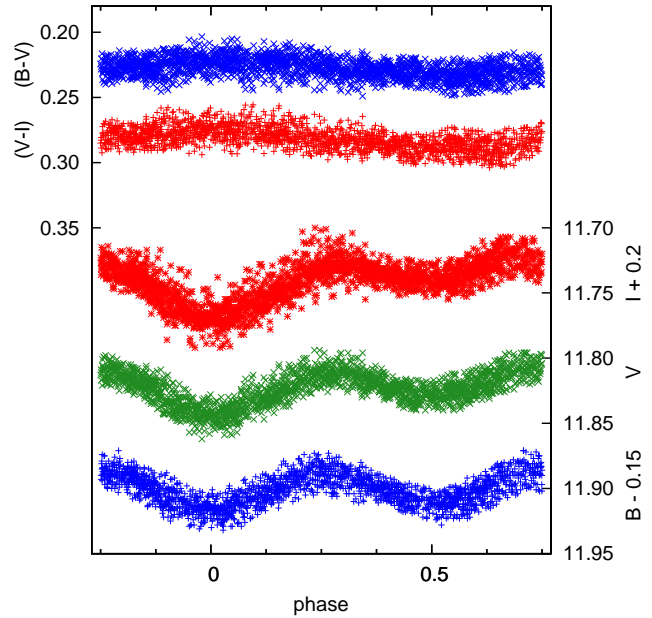}
\caption{$BVI_\mathrm{C}$ and $B-V$, $V-I_\mathrm{C}$ light and colour  curves of V1, a $\beta$~Lyr-type new variable in the field of RZ Lyr.}
\label{v1}
\end{centering}
\end{figure}

\begin{figure}
\begin{centering}
\includegraphics[width=8.6cm]{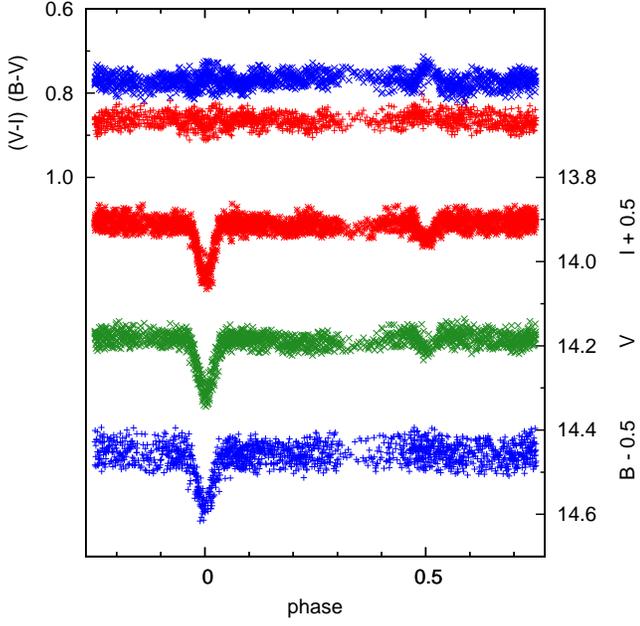}
\caption{$BVI_\mathrm{C}$ and $B-V$, $V-I_\mathrm{C}$ light and colour  curves of V2, an Algol-type new variable in the field of RZ Lyr. }
\label{v2}
\end{centering}
\end{figure}

As given in Section~\ref{konkoly}, two new variables in the field of view of RZ~Lyr have been identified. The photometric data of the new variables are given in the electronic edition of this paper (Table~\ref{newvarphot}). V1 and V2 are measured relative to the comparison stars `a' and `c', respectively (Fig.~\ref{map}). The light and colour curves these variables are shown in Figs.~\ref{v1} and \ref{v2}, respectively.
Basic information on V1 and V2 are summarized in Table~\ref{newvar}.

\begin{table*}
\caption{Basic information on the new variables}
 \label{newvar}
  \begin{tabular}{lcccccccc}
  \hline
Variable &2MASS ID         &$(V)$  &$(B-V)$&($V-I_\mathrm{C}$)&Type&Period&$T_0$[HJD]\\
\hline
V1       &18431843+3248587 &11.82&0.23   &0.28  &EB&0.927033(6)&2455353.390\\
V2       &18435377+3247459 &14.19&0.77   &0.86  &EA&1.447245(6)&2455396.370\\
\hline
\end{tabular}
\end{table*}

\begin{table}
\caption{Konkoly CCD observations of the two new variables discovered in the field of RZ Lyr. The full table is available as Supporting Information with the online version of this paper.}
 \label{newvarphot}
  \begin{tabular}{lccc}
  \hline
Variable &HJD        &mag    &Filter\\
         &2400000+   &       &      \\
\hline
V1       &55294.6409 &12.058 &$B$   \\
...      &...        &...    &...   \\
V1       &55294.6424 &11.820 &$V$   \\
...      &...        &...    &...   \\
V1       &55294.6434 &11.511 &$I_\mathrm{C}$   \\
...      &...        &...    &...   \\
V2       &55314.4615 &14.991 &$B$   \\
...      &...        &...    &...   \\
...      &...        &...    &...   \\
\hline
\end{tabular}
\end{table}



\begin{thebibliography}{}

\bibitem[Agerer et al.(2001)]{a1} Agerer F., Dahm M., Huebscher J., 2001, IBVS 5017
\bibitem[Agerer \& Huebscher(1996)]{a2} Agerer F., Huebscher J., 1996, IBVS 4382
\bibitem[Agerer \& Huebscher(2000)]{a3} Agerer F., Huebscher J., 2000, IBVS 4912
\bibitem[Agerer \& Huebscher(2003)]{a4} Agerer F., Huebscher J., 2003, IBVS 5485
\bibitem[Batyrev(1951)]{ba51} Batyrev A. A., 1951, PZ, 8, 155
\bibitem[Batyrev(1952)]{ba52} Batyrev A. A., 1952, PZ, 9, 48
\bibitem[Belik(1969)]{be69} Belik S. I., 1969, PZ, 17, 93
\bibitem[Blanco(1992)]{blanco} Blanco V. M., 1992, AJ, 104, 734
\bibitem[Bogdanov(1972)]{bo72} Bogdanov M. B., 1972, PZP, 1, 309
\bibitem[Butler et al.(1982)]{bu82} Butler D., Manduga A., Deming D., Bell R. A., 1982, AJ, 87, 640
\bibitem[\protect\citeauthoryear{Castelli \& Kurucz}{2003}]{kurucz} Castelli F., Kurucz R. L., 2003, IAUS, 210, 20
\bibitem[Chadid et al.(2010)]{ch10} Chadid M. et al., 2010, A\&A, 510, 39  %v1127 aql
\bibitem[Chadid, Gillet \& Fokin(2000)]{c00} Chadid M., Gillet, D., Fokin, A., B. 2000, A\&A, 363, 568
\bibitem[\protect\citeauthoryear{Dorman}{1992}]{dorman92}            Dorman B., 1992, ApJS, 81, 221
\bibitem[Fitch, Wisniewski \& Johnson(1966)]{fi66} Fitch W. S., Wisniewski W. Z., Johnson H. L., 1966, Comm. Lunar Plan. Lab., 5, 71
\bibitem[Guggenberger et al. (2011)]{gu11} Guggenberger E., Kolenberg K., Chapellier E., Poretti E., Szab\'o R., Benk\H{o} J. M., Papar\'o M., 2011, MNRAS, 415, 1577  %corot 105288363
\bibitem[Guggenberger et al. (2012)]{gu} Guggenberger E. et al., in preparation
\bibitem[\protect\citeauthoryear{Gillet \& Crowe}{1988}]{gc} Gillet D., Crowe R. A. 1988, A\&A, 199, 242
\bibitem[H\o g et al.(2000)]{tycho} H\o g E. et al., 2000, A\&A, 355, L27
\bibitem[Huebscher (2005)]{h1} Huebscher J., 2005, IBVS 5643
\bibitem[Huebscher (2011)]{h2} Huebscher J., 2011, IBVS 5984
\bibitem[Huebscher et al.(2005)]{h3} Huebscher J., Paschke A., Walter F., 2005, IBVS 5657
\bibitem[Huebscher et al.(2006)]{h4} Huebscher J., Paschke A., Walter F., 2006, IBVS 5731
\bibitem[Huebscher et al.(2008)]{h5} Huebscher J., Steinbach H.-M., Walter F., 2008 IBVS 5830
\bibitem[Huebscher et al.(2009)]{h6} Huebscher J., Steinbach H.-M., Walter F., 2009 IBVS 5889
\bibitem[Huebscher et al.(2010)]{h7} Huebscher J., Lehmann P.B., Monninger G., Steinbach H.-M., Walter F., 2010 IBVS 5918
\bibitem[Hurta et al.(2008)]{hu08} Hurta Zs., Jurcsik J., Szeidl B., S\'odor \'A., 2008, AJ, 135, 957 %rv uma
\bibitem[\protect\citeauthoryear{Jurcsik}{1995}]{ju95} Jurcsik J., 1995, Acta Astron., 45, 653
\bibitem[\protect\citeauthoryear{Jurcsik \& Kov\'acs}{1996}]{jk96} Jurcsik J., Kov\'acs G., 1996, A\&A, 312, 111
\bibitem[Jurcsik et al.(2008a)]{ju08} Jurcsik J. et al., 2008a, MNRAS, 391, 164	%mw lyr I.
\bibitem[Jurcsik et al.(2008b)]{mw2}  Jurcsik J. et al., 2008b, MNRAS, 393, 1553
\bibitem[Jurcsik et al.(2009a)]{dmc} Jurcsik J. et al., 2009a, MNRAS, 397, 350	%dm cyg
\bibitem[Jurcsik et al.(2009b)]{bl} Jurcsik J. et al., 2009b, MNRAS, 400, 1006	%blstat.
\bibitem[Jurcsik et al.(2011)]{m5bl} Jurcsik J., Szeidl B., Clement C., Hurta Zs., Lovas M., 2011, MNRAS, 411, 1763
\bibitem[Klepikova(1958)]{kl58} Klepikova L. A., 1958, PZ, 12, 164
\bibitem[Kolenberg et al.(2011)]{ko11} Kolenberg K. et al., 2011, MNRAS, 411, 878 %rr lyr
\bibitem[Koll\'ath(1990)]{mufran} Koll\'ath Z. 1990, Occ. Techn. Notes Konkoly Obs., No. 1, http://www.konkoly.hu/staff/kollath/mufran.html
\bibitem[Koll\'ath, Moln\'ar \& Szab\'o(2011)]{pdd} Koll\'ath Z., Moln\'ar L., Szab\'o R. 2011, MNRAS, 414, 1111
\bibitem[Le Borgne et al.(2005)]{l1} Le Borgne J.F., Klotz A., Bo\"er M., 2005, IBVS 5650
\bibitem[Le Borgne et al.(2006a)]{l2} Le Borgne J.F., Klotz A., Bo\"er M., 2006a, IBVS 5686
\bibitem[Le Borgne et al.(2006b)]{l3} Le Borgne J.F., Klotz A., Bo\"er M., 2006b, IBVS 5717
\bibitem[\protect\citeauthoryear{Le Borgne et al.}{2007a}]{lb} Le Borgne J.F. et al., 2007a, A\&A, 476, 307
\bibitem[Le Borgne et al.(2007b)]{l4} Le Borgne J.F., Klotz A., Bo\"er M., 2007b, IBVS 5767
\bibitem[Le Borgne et al.(2007c)]{l5} Le Borgne J.F., Klotz A., Bo\"er M., 2007c, IBVS 5790
\bibitem[Le Borgne et al.(2008a)]{l6} Le Borgne J.F., Klotz A., Bo\"er M., 2008a, IBVS 5823
\bibitem[Le Borgne et al.(2008b)]{l7} Le Borgne J.F., Klotz A., Bo\"er M., 2008b, IBVS 5853
\bibitem[Le Borgne et al.(2008c)]{l8} Le Borgne J.F., Vandenbroere J., Henden A.A., Butterworth N., Dvorak S., 2008c, IBVS 5854
\bibitem[Le Borgne et al.(2009a)]{l9} Le Borgne J.F., Klotz A., Bo\"er M., 2009a, IBVS 5877
\bibitem[Le Borgne et al.(2009b)]{l10} Le Borgne J.F., Klotz A., Bo\"er M., 2009b, IBVS 5895
\bibitem[Le Borgne et al.(2010)]{l11} Le Borgne J.F., Klotz A., Bo\"er M., 2010, IBVS 5934
\bibitem[Le Borgne et al.(2011)]{l12} Le Borgne J.F., Klotz A., Bo\"er M., 2011, IBVS 5986
\bibitem[\protect\citeauthoryear{Layden}{1994}]{la} Layden A., 1994, AJ, 108, 1016
 \bibitem[Migach (1969)]{m} Migach Ju. E., 1969, PZ, 16, 584
\bibitem[Poretti et al.(2010)]{po10} Poretti E. et al., 2010, A\&A, 520, 108	%corot 101128793
\bibitem[Roberts, Lehar \& Dreher(1987)]{clean} Roberts D.H., Lehar J., Dreher J.W. 1987, AL, 93, 968
\bibitem[Romanov(1969)]{ro69} Romanov Y. S., 1969, PZ, 16, 584
\bibitem[Romanov(1973)]{ro73} Romanov Y. S., 1973, AC 745, 3
\bibitem[Samolyk(2010)]{s1} Samolyk G., 2010, eJAAVSO 38, 1, 12
\bibitem[Samolyk(2011)]{s2} Samolyk G., 2011, eJAAVSO 39, 1, 23
\bibitem[\protect\citeauthoryear{Schlegel, Finkbeiner \& Davis}{1998}]{schlegel} Schlegel D. J., Finkbeiner D. P., Davis M., 1998, ApJ, 500, 525
\bibitem[\protect\citeauthoryear{Smolec et al.}{2011}]{sm} Smolec R., Moskalik P., Kolenberg K., Bryson S., Cote M. T., Morris R. L., 2011, MNRAS, 414, 2950
\bibitem[\protect\citeauthoryear{Smolec \& Moskalik}{2010}]{sm0} Smolec R., Moskalik P. 2010 A\&A 524, 40 
\bibitem[S\'odor et al.(2007)]{rrg2} S\'odor \'A., Szeidl B., Jurcsik J., 2007, A\&A, 469, 1033
\bibitem[\protect\citeauthoryear{S\'odor, Jurcsik \& Szeidl}{2008}]{ip} S\'odor \'A., Jurcsik J., Szeidl B., 2008, MNRAS, 394, 261
\bibitem[\protect\citeauthoryear{S\'odor}{2009}]{sodor_aipc}         S\'odor \'A., 2009, in Guzik J. A., Bradley P. eds, AIP Conf. Proc. 1170, Stellar Pulsation: Challenges for Theory and Observation, p. 294
\bibitem[S\'odor et al.(2011)]{czl} S\'odor \'A., et al., 2011, MNRAS, 411, 1585 %cz lac
\bibitem[Szab\'o et al.(2010)]{pd} Szab\'o R. et al., 2010, MNRAS, 409, 1244 
\bibitem[Stothers(2006)]{st1} Stothers R., 2006, ApJ, 652, 643
\bibitem[Stothers(2010)]{st2} Stothers R., 2010, PASP, 122, 536
\bibitem[Stothers(2011)]{st3} Stothers R., 2011, PASP, 123, 127
\bibitem[Sturch(1966)]{st66} Sturch C., 1966, ApJ, 143, 774
\bibitem[\protect\citeauthoryear{Suntzeff, Kraft \& Kinman}{1994}]{su} Suntzeff N. B., Kraft R. P., Kinman T. D., 1994, ApJS, 93, 271
\bibitem[Tsessevich(1953)]{ts53} Tsessevich V. P., 1953, GAIS, 23, 62
\bibitem[Tsessevich(1958)]{ts58} Tsessevich V. P., 1958, PZ, 12, 164
\bibitem[Tsessevich (1969)]{ts69} Tsessevich V. P., 1969, PZ, 16, 584
\bibitem[Williams(1903)]{wi03} Williams A. S., 1903, AN, 162, 257
\bibitem[Wisniewski \& Johnson(1968)]{bc}  Wisniewski W. Z., Johnson H. L., 1968, Comm. Lunar and Planet. Lab., 7, 57
\bibitem[Zverev \& Makarenko(1979)]{zv79} Zverev M. S., Makarenko E. N., 1979, PZP, 3, 431
\end{thebibliography}
\end{document}